\documentclass[conference]{IEEEtran}
\IEEEoverridecommandlockouts
\usepackage{cite}
\usepackage{amsmath,amssymb,amsfonts}
\usepackage{algorithmic}
\usepackage{graphicx}
\usepackage{textcomp}
\usepackage{multirow}
\usepackage{array}
\usepackage{tikz}
\makeatletter

\newcommand{\Rmnum}[1]{\expandafter\@slowromancap\romannumeral #1@}
\makeatother

\def\BibTeX{{\rm B\kern-.05em{\sc i\kern-.025em b}\kern-.08em
    T\kern-.1667em\lower.7ex\hbox{E}\kern-.125emX}}

\title{Classification of Upper Limb Movements \newline Using Convolutional Neural Network \newline with 3D Inception Block
\footnote{{\thanks{\hrule Research was partly supported by Institute of Information \& Communications Technology Planning \& Evaluation (IITP) grant funded by the Korea government (No. 2017-0-00432, Development of Non-Invasive Integrated BCI SW Platform to Control Home Appliances and External Devices by User’s Thought via AR/VR Interface) and partly funded by Institute of Information \& Communications Technology Planning \& Evaluation (IITP) grant funded by the Korea government (No. 2017-0-00451, Development of BCI based Brain and Cognitive Computing Technology for Recognizing User’s Intentions using Deep Learning).}
}
}}

\author{\IEEEauthorblockN{Do-Yeun Lee$^1$, Ji-Hoon Jeong$^1$, Kyung-Hwan Shim$^1$, Dong-Joo Kim$^1$}
\IEEEauthorblockA{{$^1$Department of Brain and Cognitive Engineering, Korea University, Seoul, Republic of Korea} \\
doyeun$\_$lee@korea.ac.kr, jh$\_$jeong@korea.ac.kr, kh$\_$shim@korea.ac.kr, dongjookim@korea.ac.kr}
}

\begin{document}

\maketitle

\begin{abstract}
A brain-machine interface (BMI) based on electroencephalography (EEG) can overcome the movement deficits for patients and real-world applications for healthy people. Ideally, the BMI system detects user movement intentions transforms them into a control signal for a robotic arm movement. In this study, we made progress toward user intention decoding and successfully classified six different reaching movements of the right arm in the movement execution (ME). Notably, we designed an experimental environment using robotic arm movement and proposed a convolutional neural network architecture (CNN) with inception block for robust classify executed movements of the same limb. As a result, we confirmed the classification accuracies of six different directions show 0.45 for the executed session. The results proved that the proposed architecture has approximately 6$\sim$13\% performance increase compared to its conventional classification models. Hence, we demonstrate the 3D inception CNN architecture to contribute to the continuous decoding of ME.\\ 
\end{abstract}

\begin{small}\textbf{\textit{Keywords-brain-machine interface; electroencephalogram; robotic arm; movement execution; deep learning}\\}\end{small}

\section{Introduction}
Brain-machine interface (BMI) has demonstrated that users can voluntarily control neuroprosthetic devices \cite{wolpaw2002brain}. Over the past decades, the BMI researches have been demonstrated as a promising technique for operating a wheelchair \cite{kim2016commanding}, a drone \cite{drone}, a robotic arm \cite{jeong2019trajectory} and etc. The BMI decodes the user's intention from the movement parameter of cortical activity. The BMI-based technique has been developed not only to rehabilitate movement functions for stroke, amyotrophic lateral sclerosis (ALS), spinal cord injury (SCI) or other patients \cite{zhu2016canonical} but also to supporting multitasks in daily life for healthy people \cite{penaloza2018bmi}. However, non-invasive BMI-based robotic arm systems have not yet appeared to successfully control the device based on upper limb movements. The robotic arm control for high-level tasks such as drinking water or moving some objects requires a complex and robust interface \cite{B2} to coordinate the high degree of freedom (DoF) of the robotic arm \cite{meng2016noninvasive}.

Most non-invasive BMI-based robotic arm systems have been designed using different types of electroencephalography (EEG) modality \cite{chen2016high,chen2017extraction,lee2017network,kim2014detection} such as a steady-state visually evoked potential (SSVEP) \cite{won2015effect, kwak2017convolutional, lee2018high}, an event-related potential (ERP) \cite{palankar2009control, yeom2014efficient,won2017motion}, error-related potential (ErrP) \cite{spuler2012online}, movement imagery (MI) \cite{suk2011subject, kam2013non} and movement execution (ME) \cite{wang2010dynamic, kim2014decoding}. In this paper, we focused on ME paradigm in which can design a robotic arm system. The ME paradigm not only generates brain activity \cite{ding2013changes, A3} for various spontaneous movements but also helps rehabilitation. Ofner et al. \cite{ofner2019attempted} conducted experiments with actual movement paradigm to SCI patients. They classified five classes of grasps and twists and showed 45\% performance. Kim et al. \cite{kim2019classification} investigated premovement before real movement. They experimented by moving the cursor over four targets and classified them for right, left, up, and down directions. 

At the same time, other conventional studies focused on accurately decoding the user’s upper limb movements and arm reaching trajectory. Korik et al. \cite{korik2018decoding} investigated the 3D hand movement trajectories from EEG using a power spectral density (PSD) based band power time-series (BTS) model. The EEG and kinematics data were acquired using the movement execution paradigm. They compared the trajectory between band-pass filtered potential time-series (PTS) model and the BTS model. They demonstrated that the low gamma (28-40 Hz) band could improve decoding accuracy for participants. Ubeda et al. \cite{ubeda2017classification} confirmed the feasibility of decoding upper limb kinematics from EEG in center-out reaching tasks during passive and active movements. They applied the multi-dimensional linear regression model with kinematics state (position and velocity). The decoding accuracy showed that the low-frequency range (0.1-2 Hz) include more significant information to continuous decoding of center-out reaching task and achieved 51.3\% classification results for four directions. Handiru et al. \cite{handiru2017eeg} applied a method called EEG source imaging (ESI) to classify four-class arm reaching movements in 2-dimensional space with improved accuracy in offline experiment compared to the sensor-based approaches.

In this paper, we proposed a convolutional neural network (CNN) architecture with inception blocks to classify six different reaching tasks for robotic arm control. Also, we constructed an experimental environment that can acquire the brain signals according to the executed session with the robotic arm.\\
\begin{figure}[t!]
\centering
\centerline{\includegraphics[width=\columnwidth]{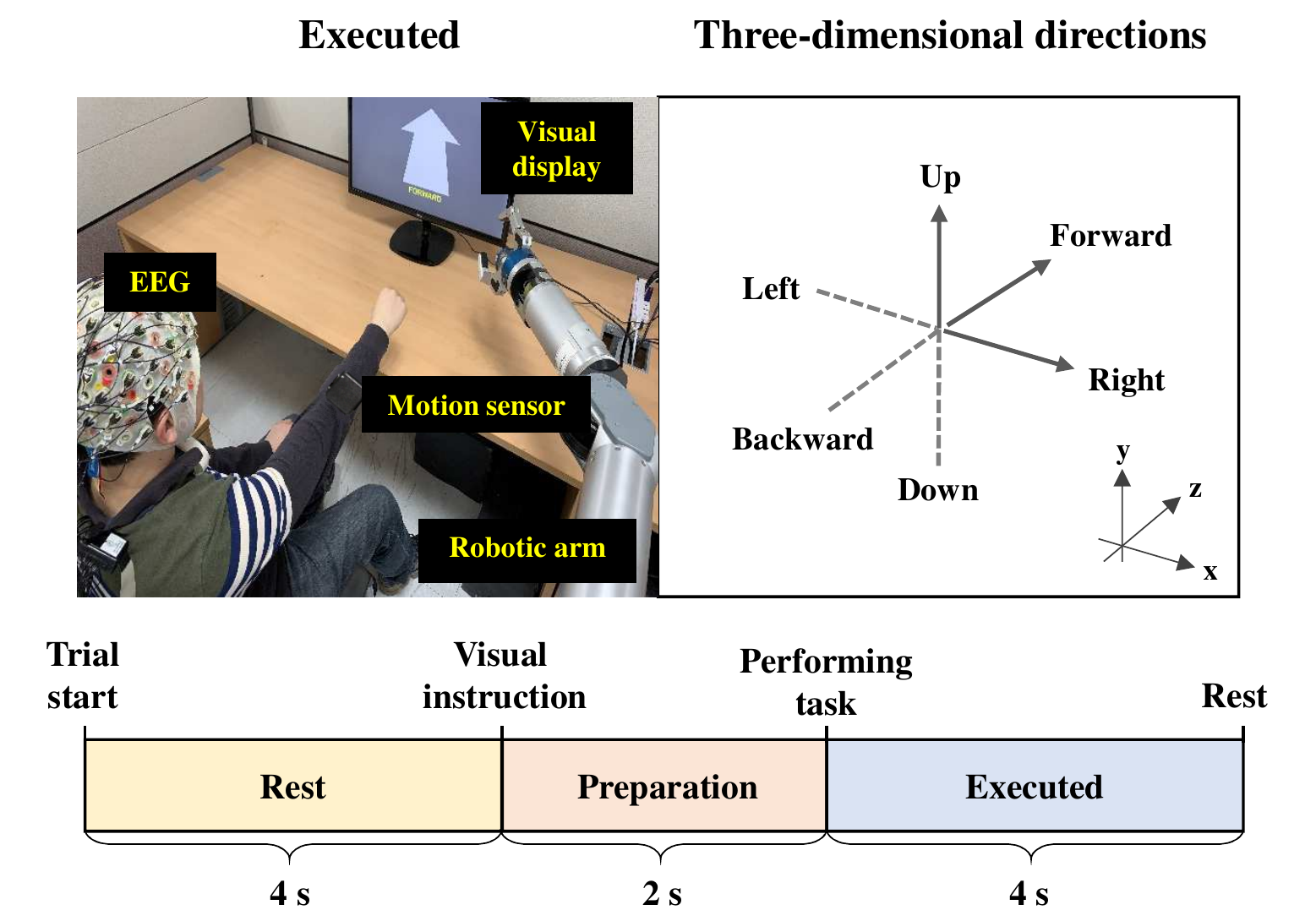}}
\caption{Experimental environments and paradigm. The participants perform six different reaching tasks through executed session according to the robotic arm movement.}
\label{fig:res}
\end{figure}   

\section{Materials and methods}

\subsection{Participants}
Five healthy and right-handed participants (all males and 23-32 years) were recruited for the experiment. None of them had prior experience with the BMI experiment. Before starting experiment, all participants were provided with an overview of the procedure. The protocols and environments were reviewed and approved by the Institutional Review Board at Korea University [1040548-KU-IRB-17-172-A-2].

\subsection{Experimental paradigm}
Participants sat comfortably near robotic arm. A visual display was placed in front so they could see the instructions. The experiment comprised of an executed session, actual movement. During the session, participants performed upper limb movements based on movement of robotic arm (Fig. 1). Participants were asked to perform the repetitive arm reaching tasks for six different directions: left, right, forward, backward, up and down. In this session, visual instructions for cue and a cross for rest were shown on the visual display, respectively (Fig. 1). We recorded 40 trials per each direction, totally 240 trials per subject. 
\begin{figure}[t!]
\centering
\centerline{\includegraphics[width=\columnwidth]{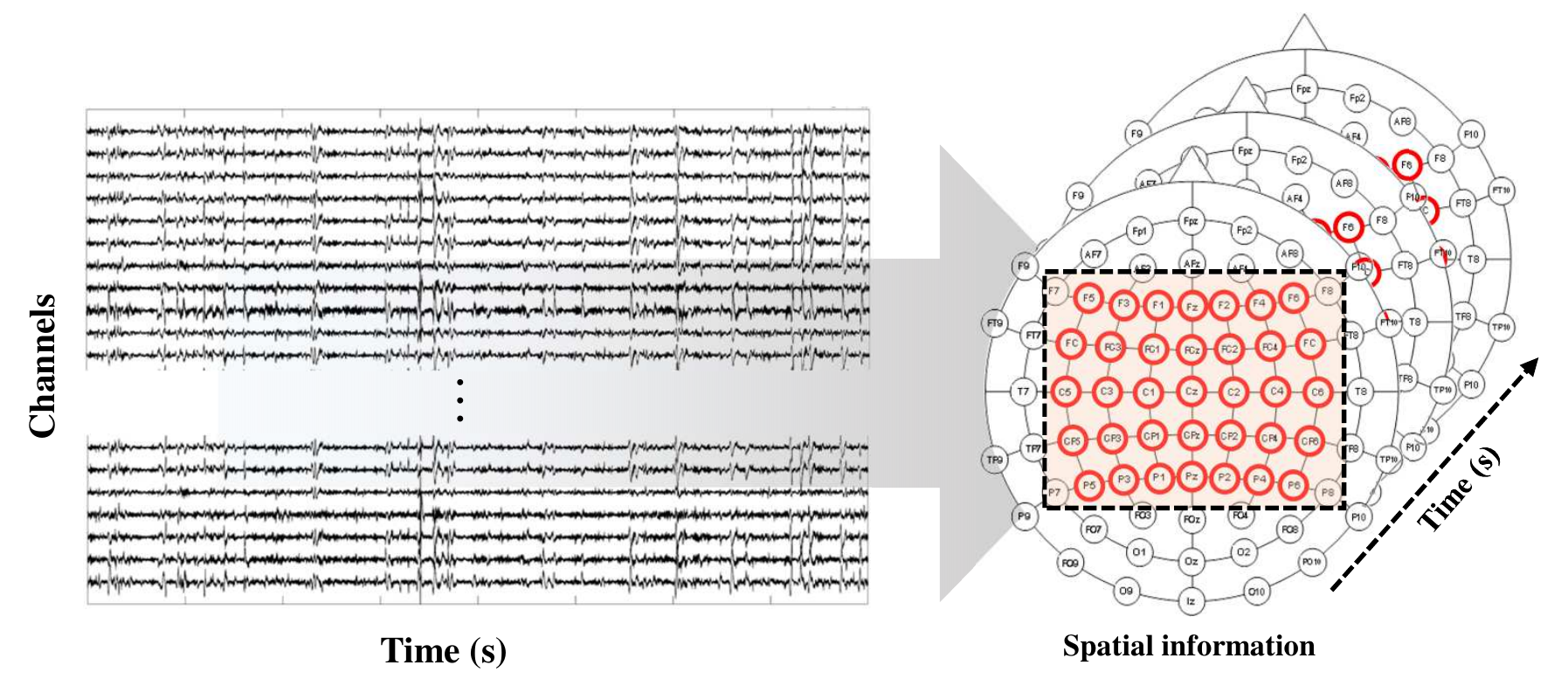}}
\caption{The transformation of the input data structure for considering three different types of brain signal's characteristics.}
\label{fig:res}
\end{figure}   

\begin{figure*}[t!]
\centering
\centerline{\includegraphics[scale = 0.7]{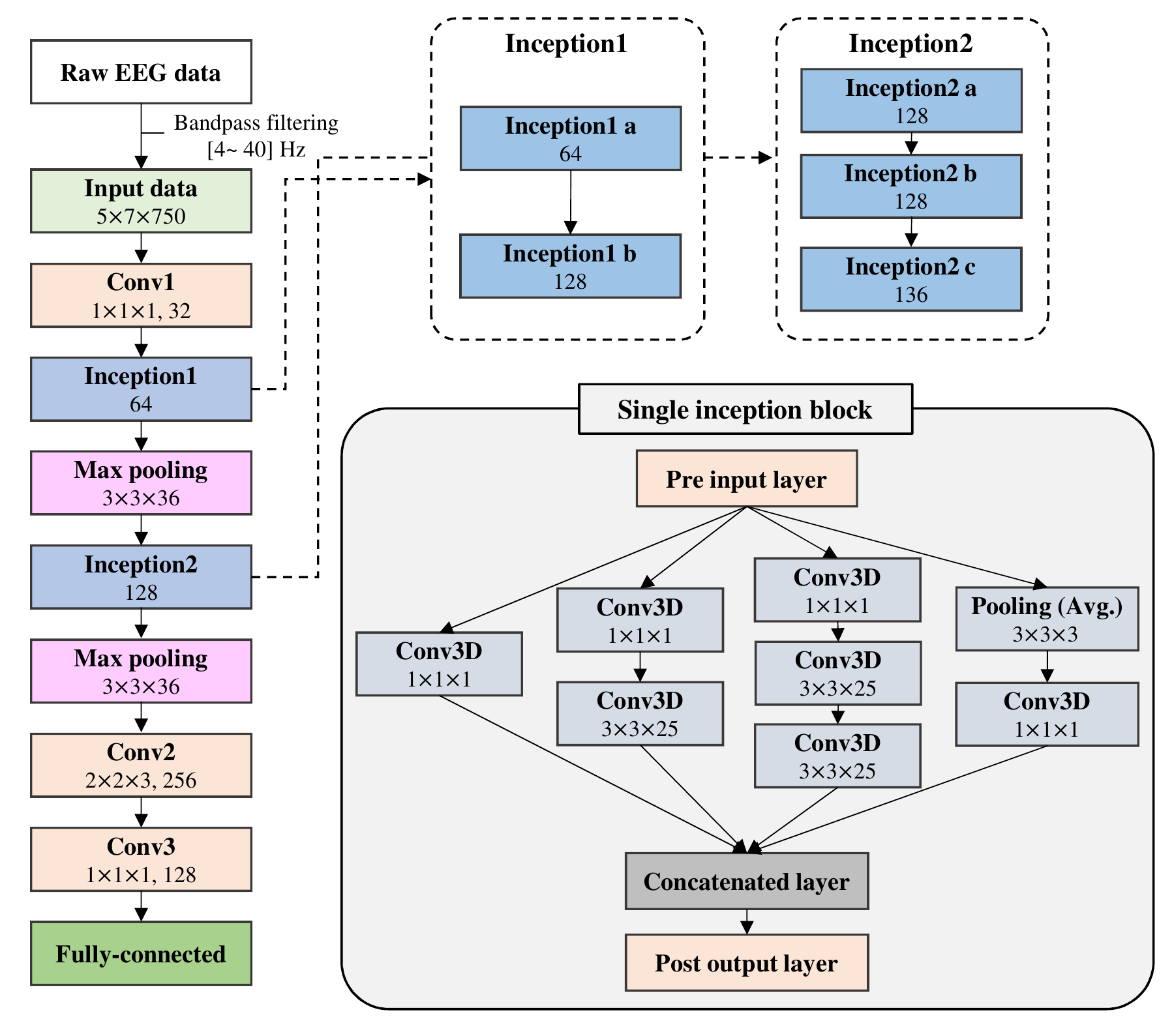}}
\caption{The flowchart of the proposed 3D inception CNN architecture. The architecture comprises convolution layers, inception block layers, pooling layers and fully-connected layer.}
\label{fig:res}
\end{figure*}

\subsection{Recording}
We recorded the EEG signals with 64 electrodes covering frontal, central and parietal areas. Reference and ground were placed on the FPz channel and FCz channel according to the international 10/20 system, respectively. Additionally, we recorded the kinematic information using motion sensor placed on the right wrist. EEG signals were recorded with ActiCap system, BrainAmp amplifiers (Brain Product GmbH, Germany) and a Matlab 2018a software. We applied an eighth-order Butterworth band-pass filter with cutoff frequencies at 0.01 and 100 Hz, a Notch filter at 60 Hz and then sampled the signals with 1000 Hz.

\subsection{Pre-processing}
The EEG signals were down-sampled from 1000 Hz to 100 Hz and were applied band-pass filtering with 4-40 Hz using zero-phase, third-order, Butterworth filter. We segmented filtered signals into training (80\% of entire data) and test data (20\% of entire data). We analyzed the 3-second data of the executed phase and selected the 20 channels (FC5, FC3, FC1, FC2, FC4, FC6, C5, C3, C1, C2, C4, C6, CP5, CP3, CP1, CPz, CP2, CP4, and CP6). Although the EEG data were obtained from 64 electrodes, data from prefrontal area were not used to avoid artifacts caused by eye movement. Also, the EEG data from the temporal and occipital areas did not use due to commonly known as a sound and visual inspection, respectively.

\subsection{Data Analysis}
The CNN has become popular recently due to their good generalization capacity and available GPU Hardware needed for parameter optimization \cite{bulthoff2003biologically}. Generally, the CNN was applied for better classification performances to various types of images. Recently, the CNN architecture was applied into the BMI fields to consider dynamics of the signal during the movement and to extract static energy feature robustly \cite{roy2019chrononet}.

In this paper, we proposed a CNN architecture with three-dimensional inception block. The architecture can contribute to consider the main characteristics of brain signals such as temporal, spectral and spatial features. The shape of input data from two-dimensions (time$\times$channels) to three-dimensions (Fig. 2) \cite{shim2019assistive}. The architecture comprises three main building layers such as convolution, inception block and max-pooling (Fig. 3). We applied the inception blocks at twice. Also, inception 1 block and inception 2 block which comprises two and three successive blocks use between the conv1 layer and max-pooling layer. Besides that, to eliminate the need of choosing the specific layer type at each level of the network, the inception blocks use four different bands of layers simultaneously. Also, the conv3D (1$\times$1$\times$1) filters are used to significantly reduce the number of network parameters by dimension reduction for feature space. The first three bands use the conv3D (1$\times$1$\times$1) filter at the beginning. In particular, the second band and third band include one conv3D (3$\times$3$\times$25) filter and two successive conv3D (3$\times$3$\times$25) filter, respectively. The fourth band of the inception block performs an average pooling operation. The number of features in each convolution depends on the input and is shown in Fig. 3. The post output layer of each inception block increases the feature dimension in 1.5 times compared to its pre-input layer. After the inception block layers, two convolution layers were used and the classification result is shown through the full-connected layer (Fig. 3).\\
\begin{table}[t!]
\small
\caption{The performance comparison of 3D inception CNN with conventional models for executed session}
\renewcommand{\arraystretch}{1,3}
\resizebox{\columnwidth}{!}{%
\begin{tabular}{cccccc}
\hline
\multirow{2}{*}{{Subject}} & \multicolumn{5}{c}{{Classification accuracy}} \\ \cline{2-6} 
 & {FBCSP} & {RF} & {\begin{tabular}[c]{@{}c@{}}Shallow\\ CNN\end{tabular}} & {\begin{tabular}[c]{@{}c@{}}3D\\ CNN\end{tabular}} & {\begin{tabular}[c]{@{}c@{}}3D Inception\\ CNN\end{tabular}} \\ \hline
{S1} & 0.30 & 0.38 & 0.50 & 0.50 & 0.51 \\
{S2} & 0.42 & 0.38 & 0.41 & 0.41 & 0.45 \\
{S3} & 0.38 & 0.39 & 0.37 & 0.37 & 0.43 \\
{S4} & 0.23 & 0.26 & 0.31 & 0.29 & 0.42 \\
{S5} & 0.26 & 0.36 & 0.35 & 0.31 & 0.31 \\
\textbf{Avg.} & \textbf{0.32} & \textbf{0.36} & \textbf{0.39} & \textbf{0.37} & \textbf{0.45} \\
\textbf{Std.} & 0.08 & 0.05 & 0.07 & 0.07 & 0.07 \\ \hline
\end{tabular}%
}
\end{table}
\section{Results and discussions}
Table \Rmnum{1} showed the classification accuracies compared with the executed session using filter-bank common spatial pattern (FBCSP), random forest (RF), shallow CNN \cite{schirrmeister2017deep}, simple 3D CNN and 3D inception CNN. There are significant differences between comparison models and ours (FBCSP and RF: \textit{p}\textless{}0.001, shallow CNN, simple CNN: \textit{p}\textless{}0.01). The 3D inception CNN had the highest grand-averaged classification accuracy according to the participants (0.45) and the FBCSP model showed the lowest results (0.32). In the executed session, the 3D inception CNN had a minimum of 6\% and a maximum of 13\% performance differences compared to the shallow CNN and the FBCSP models, respectively. 

In 3D inception CNN model, participant S1 was particularly visible who obtained the best classification accuracy for an executed session (0.51). Also there was a performance difference of about 0.2 when comparing participants S1 and S5. The maximum performance was observed to be 0.19 (participant S4) and 0.05 (participant S5). Through these results, it proves that the 3D inception CNN architecture shows a significant effect on improving decoding performance of multiclass-classification.
\begin{figure}[t!]
\centering
\centerline{\includegraphics[width=\columnwidth]{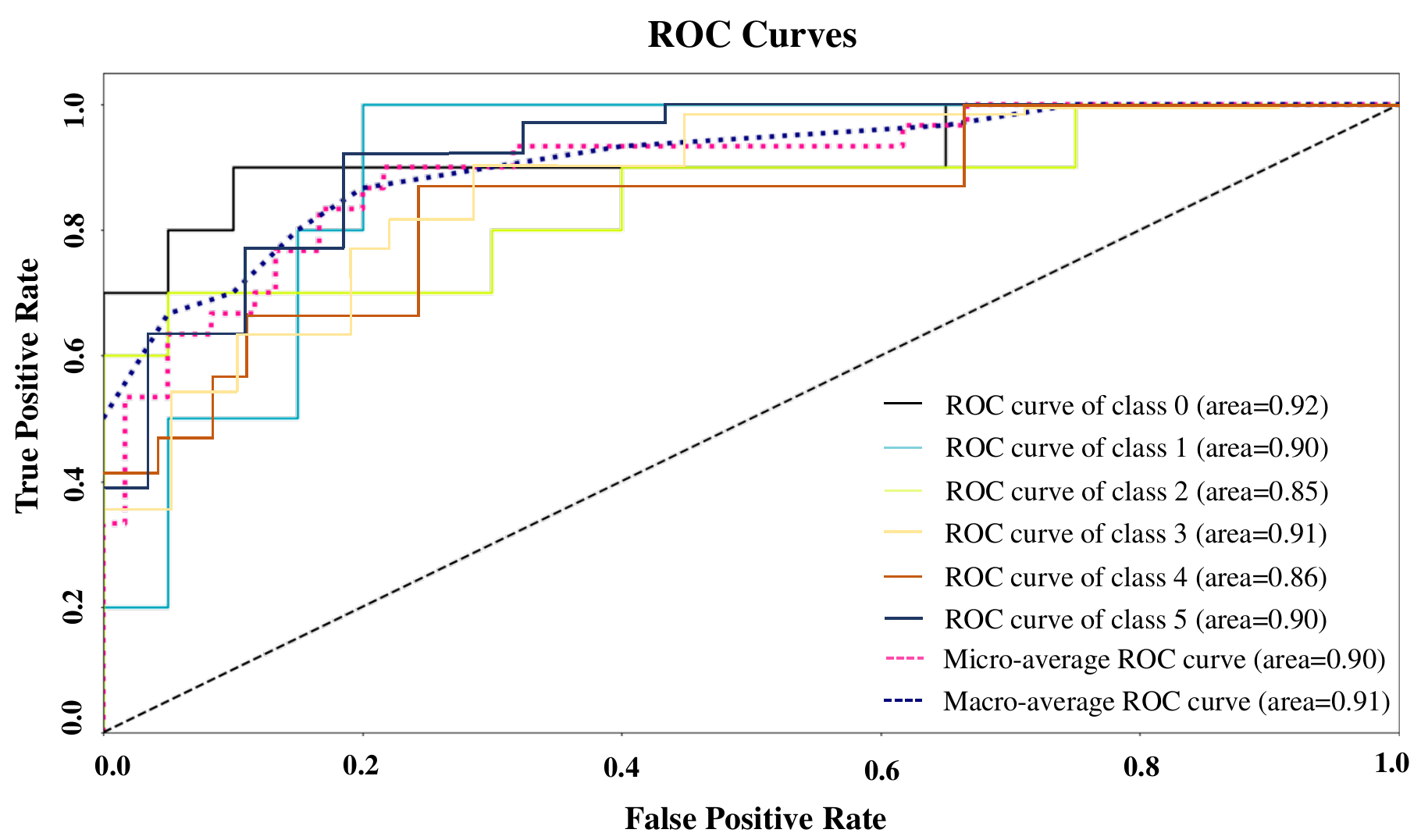}}
\caption{The ROC curves for 3D inception CNN on the six different reaching classes data.}
\label{fig:res}
\end{figure}   
\begin{figure}[t!]
\centering
\centerline{\includegraphics[width=\columnwidth]{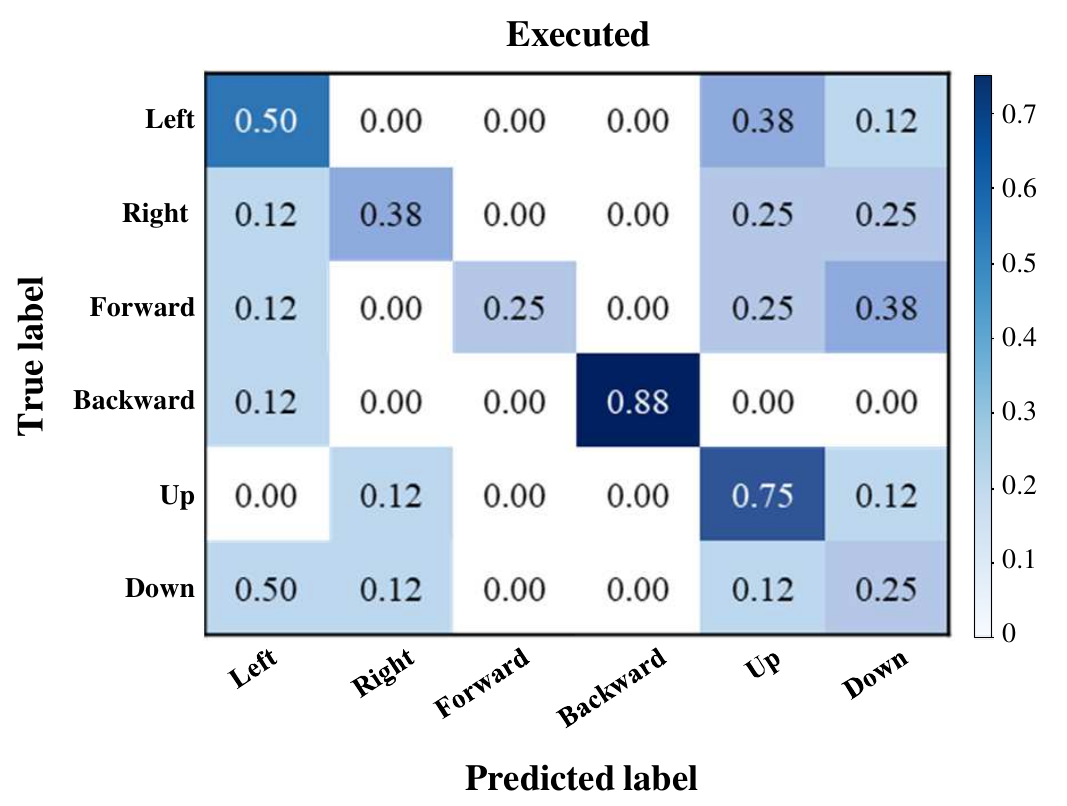}}
\caption{The confusion matrix of the six different reaching tasks for the participant S1.}
\label{fig:res}
\end{figure}   
A receiver operating characteristic (ROC) curve is presented by plotting the true positive rate (TPR) against the false positive rate (FPR) for each of the learning rate. The area under the ROC curve (AUC) means region under the curve for classifier comparison. Fig. 4 showed the ROC curves for six different reaching tasks learning rates. The learning rate of class 0 (`Backward' class) performed 0.92 and class 2 (`Down' class) had 0.85.

Fig. 5 showed a confusion matrix of a representative participant S1 according to the executed session. The matrix indicated that misclassification is mainly observed at `Forward' and `Down' classes of the 3D plane. It was a similar tendency for all participants. Participant S1 showed the highest true positive value for the `Backward' class (0.88). Also, `Forward' and `Down' classes were 0.25, the lowest values. The results showed that although a slight difference for performance between each class, there was not a large variation for classification accuracy. Thus, we have demonstrated it is possible to decode elaborate upper-limb executed movement from the robotic arm control.\\

\section{Conclusion}
In this paper, we proved that the feasibility of decoding six different reaching tasks (left, right, forward, backward, up, and down) in the three-dimensional place from EEG signals. To do that, we applied the CNN architecture based on 3D inception blocks and confirmed the robust multi-classification accuracies compared to its conventional models. The grand-averaged performance is 0.45 in an executed session that shows the higher performance difference (approximately 30\%) than chance level (16.6\%) and other models (8$\sim$13\%). Thus, we demonstrated that the 3D inception CNN model is relatively effective for decoding movement intention from EEG signals. Although BMI-based high DoF control is a difficult challenge, it is a critical issue for offering the high interaction between users and robotic arm. Our future work will be developed an asynchronous brain-controlled robotic arm system based on continuous decoding of high-level tasks.\\

\bibliographystyle{IEEEbib}
\bibliography{refs}

\begin{thebibliography}{10}

\bibitem{wolpaw2002brain}
{J. R. Wolpaw, N. Birbaumer, D. J. McFarland, G. Pfurtscheller, and T. M.
  Vaughan},
\newblock ``{Brain-computer interfaces for communication and control},''
\newblock {\em Clin. Neurophysiol.}, vol. 113, pp. 767--791, 2002.

\bibitem{kim2016commanding}
{K.-T. Kim, H.-I. Suk, and S.-W. Lee},
\newblock ``{Commanding a brain-controlled wheelchair using steady-state
  somatosensory evoked potentials},''
\newblock {\em IEEE Trans. Neural Syst. Rehabil. Eng.}, vol. 26, pp. 654--665,
  2018.

\bibitem{drone}
{K. LaFleur, K. Cassady, A. Doud, K. Shades, E. Rogin, and B. He},
\newblock ``{Quadcopter control in three-dimensional space using a noninvasive
  motor imagery-based brain-computer interface},''
\newblock {\em J. Neural. Eng.}, vol. 10, pp. 046003, 2013.

\bibitem{jeong2019trajectory}
{J.-H. Jeong, K.-H. Shim, J.-H. Cho, and S.-W. Lee},
\newblock ``{Trajectory decoding of arm reaching movement imageries for
  brain–controlled robot arm system},''
\newblock in {\em Int. Conf. Proc. IEEE Eng. Med. Biol. Soc. (EMBC)}, 2019, pp.
  23--27.

\bibitem{zhu2016canonical}
{X. Zhu, H.-I. Suk, S.-W. Lee, and D. Shen},
\newblock ``Canonical feature selection for joint regression and multi-class
  identification in alzheimer’s disease diagnosis,''
\newblock {\em Brain imaging behav.}, vol. 10, pp. 818--828, 2016.

\bibitem{penaloza2018bmi}
{C. I. Penaloza and S. Nishio},
\newblock ``{BMI control of a third arm for multitasking},''
\newblock {\em Sci. Robot.}, vol. 3, pp. eaat1228, 2018.

\bibitem{B2}
{M.-H. Lee, S. Fazli, J. Mehnert, and S.-W. Lee},
\newblock ``{Subject-dependent classification for robust idle state detection
  using multi-modal neuroimaging and data-fusion techniques in BCI},''
\newblock {\em Pattern Recognit.}, vol. 48, pp. 2725--2737, 2015.

\bibitem{meng2016noninvasive}
{J. Meng, S. Zhang, A. Bekyo, J. Olsoe, B. Baxter, and B. He},
\newblock ``{Noninvasive electroencephalogram based control of a robotic arm
  for reach and grasp tasks},''
\newblock {\em Sci. Rep.}, vol. 6, pp. 38565, 2016.

\bibitem{chen2016high}
{Y. Chen, A. D. Atnafu, I. Schlattner, W. T. Weldtsadik, M.-C. Roh, H.-J. Kim,
  S.-W. Lee, B. Blankertz, and S. Fazli},
\newblock ``{A high-security EEG-based login system with RSVP stimuli and dry
  electrodes},''
\newblock {\em IEEE Inf. Fore. Sec.}, vol. 11, pp. 2635--2647, 2016.

\bibitem{chen2017extraction}
{X. Chen, H. Zhang, L. Zhang, C. Shen, S.-W. Lee, and D. Shen},
\newblock ``{Extraction of dynamic functional connectivity from brain grey
  matter and white matter for MCI classification},''
\newblock {\em Hum. Brain Mapp.}, vol. 38, pp. 5019--5034, 2017.

\bibitem{lee2017network}
{M. Lee, R.D. Sanders, S.-K. Yeom, D.-O. Won, K.-S. Seo, H.-J. Kim, G. Tononi,
  and S.-W. Lee},
\newblock ``Network properties in transitions of consciousness during
  propofol-induced sedation,''
\newblock {\em Sci. Rep.}, vol. 7, pp. 16791, 2017.

\bibitem{kim2014detection}
{I.-H. Kim, J.-W. Kim, S. Haufe, and S.-W. Lee},
\newblock ``{Detection of braking intention in diverse situations during
  simulated driving based on EEG feature combination},''
\newblock {\em J. Neural. Eng.}, vol. 12, pp. 016001, 2014.

\bibitem{won2015effect}
{D.-O. Won, H.-J. Hwang, S. D{\"a}hne, K. R. M{\"u}ller, and S.-W. Lee},
\newblock ``{Effect of higher frequency on the classification of steady-state
  visual evoked potentials},''
\newblock {\em J. Neural Eng.}, vol. 13, pp. 016014, 2015.

\bibitem{kwak2017convolutional}
{N.-S. Kwak, K.-R. M{\"u}ller, and S.-W. Lee},
\newblock ``A convolutional neural network for steady state visual evoked
  potential classification under ambulatory environment,''
\newblock {\em PLoS One}, vol. 12, pp. e0172578, 2017.

\bibitem{lee2018high}
{M.-H. Lee, J. Williamson, D.-O. Won, S. Fazli, and S.-W. Lee},
\newblock ``{A high performance spelling system based on EEG-EOG signals with
  visual feedback},''
\newblock {\em IEEE Trans. Neural Syst. Rehabil. Eng.}, vol. 26, pp.
  1443--1459, 2018.

\bibitem{palankar2009control}
{M. Palankar, K. J. Laurentis, R. Alqasemi, E. Veras, R. Dubey, Y. Arbel, and
  E. Donchin},
\newblock ``{Control of a 9-DoF wheelchair-mounted robotic arm system using a
  P300 brain computer interface: initial experiments},''
\newblock in {\em IEEE Int. Conf. Robot. Bio.}, 2009, pp. 348--353.

\bibitem{yeom2014efficient}
{S.-K. Yeom, S. Fazli, K.-R. M{\"u}ller, and S.-W. Lee},
\newblock ``An efficient erp-based brain-computer interface using random set
  presentation and face familiarity,''
\newblock {\em PLoS One}, vol. 9, pp. e111157, 2014.

\bibitem{won2017motion}
{D.-O. Won, H.-J. Hwang, D.-M. Kim, K.-R. Müller, and S.-W. Lee},
\newblock ``Motion-based rapid serial visual presentation for gaze-independent
  brain-computer interfaces,''
\newblock {\em IEEE Trans. Neural Syst. Rehabil. Eng.}, vol. 26, pp. 334--343,
  2017.

\bibitem{spuler2012online}
{M. Sp{\"u}ler, M. Bensch, S. Kleih, W. Rosenstiel, M. Bogdan, and A.
  K{\"u}bler},
\newblock ``{Online use of error-related potentials in healthy users and people
  with severe motor impairment increases performance of a P300-BCI},''
\newblock {\em Clin. Neurophysiol.}, vol. 123, pp. 1328--1337, 2012.

\bibitem{suk2011subject}
{H.-I. Suk and S.-W. Lee},
\newblock ``Subject and class specific frequency bands selection for multiclass
  motor imagery classification,''
\newblock {\em Int. J. Imag. Syst. Tech.}, vol. 21, pp. 123--130, 2011.

\bibitem{kam2013non}
{T.-E. Kam, H.-I. Suk, and S.-W. Lee},
\newblock ``{Non-homogeneous spatial filter optimization for
  ElectroEncephaloGram (EEG)-based motor imagery classification},''
\newblock {\em Neurocomputing}, vol. 108, pp. 58--68, 2013.

\bibitem{wang2010dynamic}
{L. Wang, C. Yu, H. Chen, W. Qin, Y. He, F. Fan, Y. Zhang, M. Wang, K. Li, and
  Y. Zang},
\newblock ``{Dynamic functional reorganization of the motor execution network
  after stroke},''
\newblock {\em Brain}, vol. 133, pp. 1224--1238, 2010.

\bibitem{kim2014decoding}
{J.-H. Kim, F. Bie{\ss}mann, and S.-W. Lee},
\newblock ``Decoding three-dimensional trajectory of executed and imagined arm
  movements from electroencephalogram signals,''
\newblock {\em IEEE Trans. Neural Syst. Rehabil. Eng.}, vol. 23, pp. 867--876,
  2014.

\bibitem{ding2013changes}
{X. Ding and S.-W. Lee},
\newblock ``Changes of functional and effective connectivity in smoking
  replenishment on deprived heavy smokers: a resting-state fmri study,''
\newblock {\em PLoS One}, vol. 8, pp. e59331, 2013.

\bibitem{A3}
{M. Kim, G. Wu, Q. Wang, S.-W. Lee, and D. Shen},
\newblock ``Improved image registration by sparse patch-based deformation
  estimation,''
\newblock {\em Neuroimage}, vol. 105, pp. 257--268, 2015.

\bibitem{ofner2019attempted}
{P. Ofner, A. Schwarz, J. Pereira, D. Wyss, R. Wildburger, and G. R.
  M{\"u}ller-Putz},
\newblock ``{Attempted arm and hand movements can be decoded from low-frequency
  EEG from persons with spinal cord injury},''
\newblock {\em Sci. Rep.}, vol. 9, pp. 7134, 2019.

\bibitem{kim2019classification}
{H.-S. Kim, N. Yoshimura, and Y. Koike},
\newblock ``{Classification of movement intention using independent components
  of premovement EEG},''
\newblock {\em Front. Hum. Neurosci.}, vol. 13, pp. 63, 2019.

\bibitem{korik2018decoding}
{A. Korik, R. Sosnik, N. Siddique, and D. Coyle},
\newblock ``{Decoding imagined 3D hand movement trajectories from EEG: evidence
  to support the use of mu, beta, and low gamma oscillations},''
\newblock {\em Front. Neurosci.}, vol. 12, pp. 130, 2018.

\bibitem{ubeda2017classification}
{A. {\'U}beda, J. M. Azor{\'\i}n, R. Chavarriaga, and J. D. R. Mill{\'a}n},
\newblock ``{Classification of upper limb center-out reaching tasks by means of
  EEG-based continuous decoding techniques},''
\newblock {\em {J. Neuroeng. Rehabil.}}, vol. 14, pp. 9, 2017.

\bibitem{handiru2017eeg}
{V. S. Handiru, A. P. Vinod, and C. Guan},
\newblock ``{EEG source space analysis of the supervised factor analytic
  approach for the classification of multi-directional arm movement},''
\newblock {\em J. Neural Eng.}, vol. 14, pp. 046008, 2017.

\bibitem{bulthoff2003biologically}
{H. H. Bülthoff, S.-W. Lee, T. A. Poggio, and C. Wallraven},
\newblock ``Biologically motivated computer vision,''
\newblock {\em Springer-Verlag}, 2003.

\bibitem{roy2019chrononet}
{S. Roy, I. K. Kornek, and S. Harrer},
\newblock ``{ChronoNet: a deep recurrent neural network for abnormal EEG
  identification},''
\newblock in {\em Conf. Proc. Arti. Intel. Medi. (AIME)}, 2019, pp. 47--56.

\bibitem{shim2019assistive}
{K.-H. Shim, J.-H. Jeong, B.-H. Kwon, B.-H. Lee, and S.-W. Lee},
\newblock ``{Assistive robotic arm control based on brain-machine interface
  with vision guidance using convolution neural network},''
\newblock in {\em Conf. Proc. IEEE Int. Syst. Man Cybern. (SMC)}, 2019, pp.
  2771--2776.

\bibitem{schirrmeister2017deep}
{R. T. Schirrmeister, J. T. Springenberg, L. D. J. Fiederer, M. Glasstetter, K.
  Eggensperger, M. Tangermann, F. Hutter, W. Burgard, and T. Ball},
\newblock ``{Deep learning with convolutional neural networks for EEG decoding
  and visualization},''
\newblock {\em Human Brain Mapp.}, vol. 38, pp. 5391--5420, 2017.

\end{thebibliography}


\end{document}